\documentclass[conference]{IEEEtran}
\IEEEoverridecommandlockouts
\usepackage{cite}
\usepackage{amsmath,amssymb,amsfonts}
\usepackage{algorithmic}
\usepackage{graphicx}
\usepackage{textcomp}
\usepackage{xcolor}
\usepackage{amsmath}
\usepackage{diagbox} 
\usepackage{subcaption}
\usepackage[pass, top=0.75in]{geometry}
\usepackage[ruled,vlined]{algorithm2e} 
\usepackage{tabularx} 
\usepackage{amsmath,amssymb,bm,booktabs}
\def\BibTeX{{\rm B\kern-.05em{\sc i\kern-.025em b}\kern-.08em
    T\kern-.1667em\lower.7ex\hbox{E}\kern-.125emX}}
\begin{document}

\title{The Optimal Control Algorithm of Connected and Automated Vehicles at Roundabouts with Communication Delay\\
}

\author{\IEEEauthorblockN{1\textsuperscript{st} Chen Huang}
\IEEEauthorblockA{\textit{School of Cyber Engineering} \\
\textit{Xidian University}\\
Xi’an, China \\
}
\and
\IEEEauthorblockN{2\textsuperscript{nd} Ronghui Hou}
\IEEEauthorblockA{\textit{School of Cyber Engineering} \\
\textit{Xidian University}\\
Xi’an, China \\
}
}

\maketitle

\begin{abstract}
Connected and automated vehicles (CAVs) rely on wireless communication to exchange state information for distributed control, making communication delays a critical factor that can affect vehicle motion and degrade control performance, particularly in high-speed scenarios. To address these challenges in the complex environment of roundabout intersections, this paper proposes a roundabout control algorithm, which takes into account the uncertainty of interactive information caused by time delays. First, to maintain the required distance between the current vehicle and its preceding and following vehicles, conflicting vehicles are identified based on the time-to-collision (TTC) in the conflict zone. To fully consider communication performance, a vehicle motion model incorporating time delays is established. According to the distributed model predictive control (DMPC) mechanism, the vehicle motion control that satisfies the roundabout constraints is determined. Second, by scheduling the sequence of vehicles entering the roundabout, a multi-scale optimization objective is developed by integrating vehicle motion indicators and roundabout system indicators. Traffic density and travel time are embedded into the optimization problem to guide vehicles to enter the roundabout safely and stably. Through a variety of simulation experiments, the effectiveness of the proposed control algorithm is verified by comparing its performance with that of multiple control algorithms under different autonomous vehicle penetration rates and heavy traffic load scenarios.
\end{abstract}

\begin{IEEEkeywords}
Sequence control, Distribute model predictive control, roundabout, multi-scale optimal objectives, communication delay
\end{IEEEkeywords}

\section{INTRODUCTION}
\par With the continuous advancement of vehicle technologies, the importance of intelligent transportation systems (ITS) has significantly increased \cite{ref1}, strengthening the connection among vehicles, road networks, and users, and thereby improving the overall safety and operational efficiency of the system \cite{ref3}\cite{ref4}. Roundabouts, as a common type of intersection, offer the major advantage of maintaining smooth traffic flow while reducing traffic accidents by 50\% to 70\% \cite{ref5}, providing superior traffic efficiency and safety performance \cite{ref6,ref7,ref8,ref9,ref10}. Since roundabouts operate without traffic signals, vehicles must make autonomous decisions about when to enter, based on their own capabilities, while maintaining safe distances from surrounding vehicles. Consequently, the emergence of connected and automated vehicles (CAVs) presents a valuable opportunity \cite{ref11}. Equipped with vehicle-to-everything (V2X) communication devices, CAVs exchange data through wireless networks to achieve real-time information sharing, thereby enhancing environmental perception. Within and around roundabout areas, automated vehicles share their states and operational information via vehicle communication to ensure safe and efficient traffic operations \cite{ref12,ref13}.
\par Although extensive research has been carried out on vehicle operations within roundabouts~\cite{ref15,ref18,ref24,ref25,ref26}, most of these studies fail to account for how communication performance—especially communication delay and information acquisition processes—affects the stability of mixed traffic flow. 
Meanwhile, with the surge in the number of vehicles, a massive amount of data is generated in the vehicular network, leading to a significant increase in communication delays for all entering vehicles during peak hours. This severely impacts the throughput of roundabouts and increases safety risks~\cite{ref14}. Communication delays further exacerbate motion uncertainty among vehicles~\cite{ref20}. Moreover, due to high node mobility, communication efficiency may deteriorate even further~\cite{ref21}. The uncertainty of interactive information caused by communication delays poses new challenges for the design of control algorithms.
\par To address the challenges of conflict resolution, trajectory planning, and distributed vehicle control in multi-intersection roundabout scenarios, inspired by research in \cite{ref15} and considering communication delays in high-speed vehicle environments, we propose a robust control algorithm. An upper-level control algorithm is introduced, which determines the optimal insertion position and entry time for vehicles approaching from each direction based on a multi-scale optimization objective. Subsequently, a lower-level control algorithm is designed: within the roundabout, the distributed model predictive control (DMPC) takes into account the impact of communication latency on control performance, predicts the system states that fail to be interacted successfully, and guides each vehicle to execute optimal motion control by utilizing the states of its preceding and following vehicles. The proposed solution contributes to the realization of cooperative scheduling at intersections, enhances the vehicles' situational awareness capability, and mitigates the oscillation phenomenon among vehicles within the roundabout.
\par The novel contributions of this paper are as follows: vehicles are guided to enter the roundabout in an optimal sequence based on real-time traffic conditions and the motion states of the leading and following vehicles near each conflict zone, ensuring overall traffic efficiency. The system explicitly considers the impact of communication delays within the vehicle interaction range and identifies the preceding and following vehicles based on vehicle states and control inputs. Furthermore, constraints are categorized and scaled to enhance control robustness, ensuring the continuous safe operation of vehicles throughout the roundabout.

\section{PRELIMINARIES}
\subsection{System Architecture}
\par In the system, the considered scenario is a signal-free roundabout with dual lanes {$l_1$-$l_6$} at each entrance, as shown in Fig. 1 \cite{ref15}. Connected and automated vehicles (CAVs) share inter-vehicle information to achieve cooperative control within the roundabout. Each entrance is divided into four zones: the ramp merging zone, the conflict coordination zone, the circulating zone, and the exiting zone. The system includes merging points (MPs) ($M_1$–$M_3$), which serve as potential conflict zones for vehicles coming from different directions. Each merging point computes the optimal entry position for incoming vehicles using local control rules and shared communication data. Vehicles autonomously determine the best merging instant based on their dynamic states and surrounding traffic conditions. This mechanism ensures cooperative interaction at all entry and exit points, enhancing flow stability and intersection throughput.
\begin{figure}[!t]
    \centering
    \includegraphics[width=0.45\textwidth]{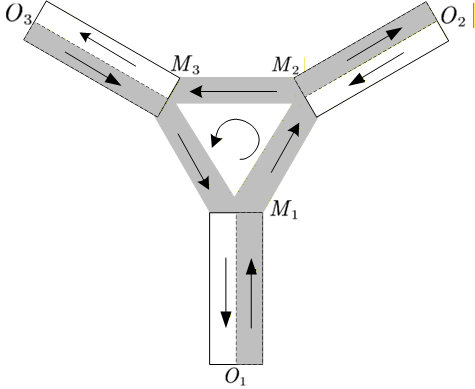}
    \caption{Performance comparison of roundabout control algorithms}
    \label{fig:roundabout_perf}
\end{figure}
\subsection{Graph Structure of Communication Topology}
\par For a group of vehicles in a platoon, the vehicles communicate and interact with the preceding vehicle and the leading vehicle via V2X devices. We denote a graph $G=\left\{ \mathcal{N},\mathcal{E} \right\} $ consisting of nodes $\mathcal{N}=\left\{ 1,2,....,N \right\} $ and edges $\mathcal{E}\subseteq \mathcal{N}\times \mathcal{N}$, where $\left( i,j \right) \in \mathcal{E}$ captures the existence of a link from node $i$ to node $j$\cite{ref27}. A graph is connected if an undirected path exists between all pairs of nodes, where each edge $(i,j)$ is associated with a non-negative weight $w_{ij}\geq 0$. Vehicles exchange information with surrounding vehicles over communication links. 
\subsection{Position Mapping}
\par For each merging point (MP), vehicles on both the ramp and the roundabout must perform position confirmation, allowing the incoming vehicle to identify its preceding and following vehicles. A position-mapping mechanism is employed, where the sequence of multiple vehicles is determined based on a virtual $Z$-axis. The specific position-mapping process is illustrated in Fig.~2 \cite{ref19}. Based on the position mapping, each vehicle identifies its preceding and following vehicles, thereby determining the optimal timing for entering the roundabout.
\begin{figure}[htbp]
    \centering
    \includegraphics[width=0.35\textwidth]{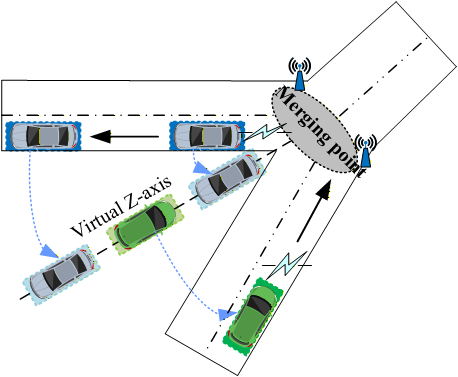} 
    \caption{ The position mapping of each merging point.}   
    \label{fig:example1}            
\end{figure}

\section{PROBLEM FORMULATION}
\par The control algorithm integrates sequencing planning with motion control. Its primary objective is to develop a roundabout control scheme that ensures traffic flow and driving safety in the presence of communication delays. The proposed control algorithm minimizes the impact of communication delays on control by evaluating and quantifying both system states and the surrounding environment. This algorithm dynamically adjusts the control inputs of each vehicle, guiding them to enter the roundabout smoothly, while ensuring coordinated and consistent driving behavior based on the positions of the preceding and following vehicles.
\par In this section, the control algorithm is developed using distributed control techniques, enabling each vehicle to follow safe and efficient trajectories based on real-time changes in the positions of its preceding and following vehicles. This algorithm ensures that vehicles can make optimal control decisions under communication delays by leveraging interactive state information. Next, we design the control optimal problem, which determines the optimal vehicle sequence based on multi-scale objectives. By using travel time as a metric, the optimal control input quantifies the overall traffic performance of the system and establishes the optimal vehicle order within each conflict zone.

\subsection{The Distribute Controller of Vehicle}
\subsubsection{Vehicle Dynamic Model with Communication delay}
\par In the roundabout intersection, vehicles use the control mechanism to plan their trajectory, velocity, and acceleration. The controller is developed to address communication delays under poor communication conditions. This ensures that each vehicle can autonomously determine its driving behavior and safely travel along the optimal path from the entrance to the exit.
\par Each vehicle in the roundabout follows a predefined path along the lane trajectory, with its trajectory curve designed to be smooth at all times. The vehicle dynamics are described using the bicycle model \cite{ref15}, which can be expressed by the following equations:
\begin{equation}
\begin{aligned}
x_i(k\!+\!1) &= x_i(k) + \Delta t \, [v_i(k)\cos\theta_i(k)],\\
y_i(k\!+\!1) &= y_i(k) + \Delta t \, [v_i(k)\sin\theta_i(k)],\\
\theta_i(k\!+\!1) &= \theta_i(k) + \Delta t \left[\frac{v_i(k)}{L_i}\tan\big(\vartheta (k-\tau)\big) \right],\\
v_i(k\!+\!1) &= v_i(k) + \Delta t \, [a_i(k-\tau) ].
\end{aligned}
\label{eq:discrete_vehicle_dynamics}
\end{equation}
where $x_i(k)$ and $y_i(k)$ represent the positions along the coordinate axes of vehicle $i$ at timestep $k$, respectively. $\theta_i(k)$ is the heading angle of vehicle $i$, $v_i(k)$ is the longitudinal velocity of vehicle $i$, $\vartheta_i(k)$ is the front steering angle of vehicle $i$, $a_i(k)$  is the longitudinal acceleration input of vehicle $i$, $L_i$ is the wheelbase length of vehicle $i$. $\Delta t$ denotes the sampling period. $\tau$ represents the communication delay of the system.
\par We define the state $s_i(k)$ and input $u_i(k-\tau)$ as
\begin{equation}
s_i(k) =
\begin{bmatrix}
x_i(k) \\[3pt]
y_i(k) \\[3pt]
\theta_i(k) \\[3pt]
v_i(k)
\end{bmatrix}, 
u_i(k-\tau) =
\begin{bmatrix}
a_i(k-\tau) \\[3pt]
\vartheta_i(k-\tau)
\end{bmatrix}, 
\end{equation}
\par The equation of system state is expressed as below. 
\begin{equation}
s_i(k+1) =
Fs_i(k) + Gu_i(k),
\end{equation}
where the matrices $F$ and $G$ are determined according to (3). 
\par Considering the communication delay of the roundabouts, vehicles obtain environmental awareness through interactive information. However, when the communication delay becomes excessively large, the vehicle cannot update its real-time control input in time. We define the communication delay threshold $T_{\mathrm{th}}$ and let the actual delay at time $k$ be $k_D(k)$. Then we introduce the indicator to represent whether the vehicle update the information of surrounding vehicles or not:
\begin{equation}
\delta(k)=
\begin{cases}
1, & k_D(k)\le T_{\mathrm{th}},\\[4pt]
0, & k_D(k)> T_{\mathrm{th}}.
\end{cases}
\end{equation}
\par When the surrounding-vehicles' information arrives within the threshold, the controller obtains the current state $s_i(k)$ and applies the nominal control law. If information for time $k$ is unavailable, the controller uses the most recent known control $u_i(k-1)$ and predicts the state. We use the prediction of system state $\tilde s(k)$ by one-step prediction method:
\begin{equation}
    \tilde s_i(k+1)=F s_i(k) + G u_i(k-1).
\end{equation}
\par
Thus, the state $\hat{s}_i(k)$ of vehicle $i$ with communication used by the controller can be written as:
\begin{equation}
    \hat{s}_i(k)=\delta(k)\,s_i(k) + (1-\delta(k))\,\tilde s_i(k),
\end{equation}
\par In this case, the corresponding control input $\hat{u}_i(k)$ of vehicle $i$ is updated as follows:
\begin{equation}
    \hat{u}_i(k)=\delta(k)\,u_i(k) + (1-\delta(k))\,u_i(k-1).
\end{equation}
\subsubsection{The Problem of Distribute Model Predictive Control}
\par To minimize the control cost $J$ while simultaneously minimizing the communication cost with communication delay. The control optimization objective is formulated as follows \cite{ref16}:
\begin{equation}\label{eq:mpc_with_delay_penalty}
\begin{aligned}
&\min_{\hat{u}_{i}(k)} \; J_i(\hat{u}_{i}(k)) 
=\; \sum_{k=t}^{t+N_p-1} \Big[ (\hat{s}_i(k) - s^{ref}_i)^T R (\hat{s}_i(k) - s^{ref}_i)
\\& \qquad +\; \hat{u}_i(k)^T Q \hat{u}_i(k) \Big]
\;+\; \lambda \sum_{k=t}^{t+N_p-1} \bar{\tau}_i .
\end{aligned}
\end{equation}
where $R$ and $Q$ are positive definite weighting matrices, $N_p$ is the prediction horizon. $s^{\mathrm{ref}}_i(k)$ denotes the reference value of state. $\sum_{k=t}^{t+N_p-1}\bar{\tau}_i$ represents the penalty term based on the communication delay and is optimized based on the following sections. $\bar{\tau}_i$ denotes the average transmission delay of the system. $\lambda$ represents the weighting coefficient corresponding to the compensation cost.
\par Note that in a roundabout with multiple entrances, the number of vehicles is typically large. Optimizing the communication routing paths among vehicles can significantly improve the overall communication performance. Reasonably scheduling data-transmission paths to achieve load balancing is an effective approach to improving communication performance. By incorporating various next-hop selection metrics, the transmission delay within the system can be effectively reduced, which further enhances the control performance in the optimization process\cite{ref23}. 
\subsubsection{Constraints}
\par The constraint conditions of vehicles within the roundabout are closely related to their positions. We define the Time-to-Collision (TTC) to describe which potential ego vehicles may pose a collision risk to vehicle $i$ in the motion process\cite{ref15}. The expression of TTC is as follows:
\begin{equation}
TTC = \frac{\text{d}_{ego,M} - \text{d}_{i,M}}{\bar{v}_{ego} - v_{i}}
\label{eq:ttc}
\end{equation}
where $\text{d}_{ego,M}$ is  the distance from potential ego vehicle to the Merge Point and $\text{d}_{i,M}$ is the distance from vehicle $i$ to the Merge Point. $\bar{v}_{ego}$ is the velocity of the ego vehicle and the value is obtained based on communication interactions. Thus $\bar{v}_{ego}(k)=v_{ego}(k-\tau)$. 
\par We denote $TTC_{\text{th}}$ to represent the threshold of collision risk. A vehicle is considered to be in a high-risk state when the TTC is smaller than the predefined threshold. $TTC_{\text{th}}$ is typically set to $2s-3s$ based on safety requirements. Only when the vehicle computes the TTC, it also indicates that the vehicle is entering the conflict area. 




\par There are these constraints into the following types: physical limitations of vehicles, positional and temporal constraints within the circular intersection, and limitations on vehicle position, velocity, and acceleration along the circular lanes \cite{ref28}. 
\begin{subequations}
\begin{equation}
TR - \frac{w_d}{2} \le \|x_i(k) - or\| \le TR + \frac{w_d}{2}, 
\quad \forall i \in \mathcal{N}, \; k \in [0, K]
\end{equation}
\begin{equation}
D_{i,i_p}(k)=x_{i_p}(k) - x_i(k) \ge \varsigma  \cdot v_i(k) + \varrho, \quad \forall i \in \mathcal{N}
\end{equation}
\begin{equation}
D_{i,i_m}(k)=x_i(k) -x_{i_m}(k)  \ge \varsigma \cdot v_i(k) + \varrho, 
\quad \forall i \in \mathcal{N}
\end{equation}
\begin{equation}
\epsilon_i(x_i(k)) \cdot v_i^2(k) \cdot hc \le w_{hc} \cdot g, 
\quad \forall i \in \mathcal{N}, \; \forall k \in [0, K]
\end{equation}
\begin{equation}
v_{\min} \le v_i(k) \le v_{\max}, \quad|a_i(k)| \le a_{\max}, 
\quad \forall i \in \mathcal{N}, \; k \in [0, K]
\end{equation}
\begin{equation}
\rho(k) \le \rho_{\max}, 
\quad k \in [0, K]
\end{equation}
\end{subequations}
where $K$ is the total time. $or$ is the center of the roundabout, $TR$ and $w_d$ are the radius and lane width. 
$\rho(k)$ denotes the vehicle density on the roundabout.$ \epsilon_i(x_i(k))$ is the curvature of the road at position $x_i(k)$ along the trajectory of vehicle $i$. $x_{i_p}(k)$ and $x_{i_m}(k)$ represent the position of the proceeding vehicle and the following vehicle, respectively. $hc$ is the height of the center of gravity of the vehicle and $w_{hc}$ is the Half-width of the vehicle. $g$ is the Gravitational acceleration constant. $\varsigma$ is reaction time (typically $\varsigma= 1.8\,\text{s}$ is recommended). $\varrho$ is related to the vehicle length $L$. $v_{\min}$, $v_{\max}$ and $a_{\max}$ denote the minimum speed, maximum speed, and maximum acceleration of the vehicle, respectively.
\par When both the proceeding vehicle $i_p$ and the following vehicle $i_m$ exist, the vehicle determines its position according to its index number, and the constraints (10b),(10c) are applied accordingly. Meanwhile, considering the influence of communication delay on inter-vehicle spacing, and in order to ensure the safety of vehicle operations, the spacing-related constraints are expressed as follows:
\begin{subequations}\label{eq:delay_constraints}
\begin{align}
D_{i,i_p}(k) \ge\; & D_{\min} + v_i(k)\,T_h  
+ v_i(k)\,\bar{\tau}  
+ \tfrac{1}{2}\,a_{\max}\,\bar{\tau}^2 ,
\tag{\theequation a} \label{eq:delay_constraints_a}
\\[6pt]
D_{i,i_p}(k) \ge\; & L + \varrho + v_i(k)\,\bar{\tau},
\tag{\theequation b} \label{eq:delay_constraints_b}
\\[6pt]
D_{i_m,i}(k) \ge\; & D_{\min} + v_{i_m}(k)\,T_h 
+ v_{i_m}(k)\,\bar{\tau},
\tag{\theequation c} \label{eq:delay_constraints_c}
\end{align}
\end{subequations}
where $D_{i,i_p}(k)$ and $D_{i_m,i}(k)$ represent the distance between vehicle $i$ the proceeding vehicle and the following vehicle. $T_h$ denotes the time headway. $D_{\min}$ is the safe spacing distance. 

\par In the DMPC-based optimal vehicle control problem (8), the constraints are transformed into a differentiable robustness function, which is incorporated as a penalty term into the optimization problem. An NLP solver is employed to compute the optimal control inputs ${U_0, U_1, \ldots, U_{N-1}}$ for the DMPC optimization problem, where the optimal acceleration and steering decisions guide each vehicle to consistently maintain a safe distance from its preceding vehicle.

\subsection{The Controller of Intersections}
\subsubsection{Sequence Control}
\par To optimize the timing at which vehicles enter the roundabout, we construct virtual platoon at each merging point by organizing both circulating vehicles on the roundabout and approaching vehicles at the entry\cite{ref19}. 
We denote the set of vehicles by $\mathcal{N}$ and the set of roundabout segments by $\mathcal{V}$, where $N = |\mathcal{N}|$ represents the total number of vehicles and $V = |\mathcal{V}|$ denotes the total number of roundabout segments. For the entire roundabout system, vehicles enter the circulating roadway according to the control algorithm, while their accelerations are determined by the distribute controller. The joint control inputs consist of the vehicle sequence $\{C_0, C_1, \ldots, C_N\}$ for each merging point and the control inputs $\{U_0, U_1, \ldots, U_{N-1}\}$ of the vehicles.
The sequence control determines the merging order between any CAV entering from an arbitrary inlet and the vehicles already traveling inside the roundabout. The binary decision matrix $C$ is defined as follows.
\begin{equation}
C(k) =
\begin{bmatrix}
c_{1,1}(k-\tau) & c_{1,2}(k-\tau) & \cdots & c_{1,N}(k-\tau) \\
c_{2,1}(k-\tau) & c_{2,2}(k-\tau) & \cdots & c_{2,N}(k-\tau) \\
\vdots & \vdots & \ddots & \vdots \\
c_{N,1}(k-\tau) & c_{N,2}(k-\tau) & \cdots & c_{N,N}(k-\tau)
\end{bmatrix}
\label{eq:U_definition}
\end{equation}
where $c_{i,n} \in \{0,1\}$ indicates whether vehicle $i$ occupies position $n$ in the virtual car-following (CF) sequence. The element $c_{i,n}$ takes on a value of 1 if the $i_{th}$ CAV on real roads is assigned with ID $n$ in the virtual CF sequence, and 0 otherwise. When the vehicle enters the ring road is related to the vehicle states on the lanes. 
Then the corresponding sequence constraints are as follows \cite{ref19}:
\begin{subequations}\label{eq:assignment}
\begin{align}
    &\sum_{i=1}^{N} c_{i,n} = 1, \sum_{n=1}^{N} c_{i,n} = 1,\quad i,n = 1,\dots,N \label{eq:assignment:b}\\
    &c_{i,n} \le 1 - \sum_{l=n+1}^{N} c_{i-1,l}, \quad i=2,\dots,N \label{eq:assignment:c}
\end{align}
\end{subequations}
\par Meanwhile, for each merge point, the control objective of sequence control is to minimize the control cost. We define $\varDelta d_i$ as the deviation between the current vehicle position and the target position of vehicle $i$. $\varDelta d_i=\left(c_{:,i}^T \hat{p}^T - c_{:,i+1}^T \hat{p}^T - \varDelta d_{i+1}^*\right)$, $\varDelta d_{i+1}^*$ represents the constant desired spacing from the predecessor of the $(i+1)_{th}$ CAV in the virtual CF sequence. $c_{:,i}$ represents the $i_{th}$ column of matrix $C$\cite{ref19}. $\hat{p}_i$ represents the current position of the vehicle $i$ based on $x_i$ and $y_i$ at the timestep $k$. $\hat{p}_i(k)=[x_i(k-\tau),y_i(k-\tau)]$. The indicator $Y_{1,n}$ and $Y_{2,n}$ are introduced to represent the relationship between the spacing deviation variable and its desired value. The binary variables $Y_{1,n}$ and $Y_{2,n}$ are introduced and the value set is [0,1]. For $n \in {1,2,\ldots,N}$:
\begin{equation}
Y_{\hat{p},n} =
\begin{cases}
Y_{1,n}, & \text{if } \left(\varDelta d_i - \varDelta d^{*}_{n+1}\right) < 0,\\[6pt]
Y_{2,n}, & \text{if } \left(\varDelta d_i- \varDelta d^{*}_{n+1}\right) > 0,
\end{cases}
\label{eq:Y_definition}
\end{equation}
\par Then the position deviation between vehicle $i$ and vehicle $n$ is denoted as $Y_{\Delta d,n}= Y_{1,n}-Y_{2,n}$. 
Similarly, we denote the velocity deviation as $Y_{\hat{v},n}$ and we also define $Y_{3,n}$ and $Y_{4,n}$ as the indicator to represent the relationship between the velocity deviation variable $\hat{v}$. We denote $\hat{v}_i(k)=v_i(k-\tau)$ to represent the velocity of vehicle $i$ at timestep $k$.  The specific expressions are given below for $n \in \{1,2,\ldots,N\}$:
\begin{equation}
Y_{\hat{v},n} =
\begin{cases}
Y_{3,n}, & \text{if } \left(c^{T}_{:,n+1}\hat{v}^{T} - c^{T}_{:,n}\hat{v}^{T}\right) < 0,\\[6pt]
Y_{4,n}, & \text{if } \left(c^{T}_{:,n+1}\hat{v}^{T} - c^{T}_{:,n}\hat{v}^{T}\right) > 0,
\end{cases}
\label{eq:Yv_definition}
\end{equation}
\par Meanwhile, $Y_{\hat{v}_{in}}$ is denoted as $Y_{\hat{v},n} = Y_{3,n}-Y_{4,n}$. 
\par Finally, we define $\chi_{in}=Y_{\Delta d,n}- Y_{\hat{v},n}$ as an index to describe how the velocity and position deviation variables change when the position sequence $C$ varies. The second optimal objective of sequence control is to minimize $\chi_{in}$ helps to reduce the spacing deviation \cite{ref19}.
\subsubsection{Travel Time}
\par Considering these merging points in roundabout, the state of each vehicle is influenced not only by the conditions at its current merging point but also by the vehicle states at other merging points. Therefore, we define a macroscopic objective to characterize the overall traffic performance of the entire roundabout. To quantitatively characterize the traffic performance of the roundabout, we define travel time as the metrics to describe the current traffic conditions within the roundabout. The entry and exit times of vehicle $i$ as $t_{i}^{\text{entry}}$ and $t_{i}^{\text{exit}}$, respectively. 
\par Vehicles need to wait for the optimal timing to enter the loop according to the optimal position decision matrix. We define the waiting delay as $t^{\text{delay}}$. At this stage, the travel time is divided into three phases according to the vehicle’s position: from the entry to the merging point, the waiting period at the merging point, and from the merging point to the exit. The time stage are denoted as $t_{i}^{\text{entry-merge}}$, $t_{i}^{\text{delay}}$  and $t_{i}^{\text{merge-exit}}$. The travel time $t_{i}^{\text{travel}}$ of vehicle $i$ is then calculated as follows:
\begin{equation}
t_{i}^{\text{travel}}=\left( t_{i}^{\text{exit}} - t_{i}^{\text{entry}} \right)=t_{i}^{\text{entry-merge}} + t_{i}^{\text{delay}} + t_{i}^{\text{merge-exit}},
\label{eq:safety_constraint1}
\end{equation}
\subsubsection{Traffic Density}
\par Traffic density is also the integrated indicators for evaluating the system performance. Based on the vehicles currently traveling within the roundabout, the traffic density $\rho_j$ of intersection $j$ is represented as 
\begin{equation}
\rho_j = \frac{N_j}{L_{\mathcal{V}_j}}, \quad \bar{\rho} = \frac{\sum_j N_j}{\sum_j L_{\mathcal{V}_j}}.
\label{eq:density}
\end{equation}
where $N_j$ represents the total number of vehicles in the roundabout. $L_{\mathcal{V}_j}$ denotes the length of the roundabout. $\bar{\rho}$ represents the average density of the system.
\begin{equation}
N_j = N_j^0 + \sum_{i=1}^{V} e_{i,j}, \quad e_{i,j} \in \{0,1\}, \forall i,j.
\label{eq:Ns-def}
\end{equation}
where $N_j^0$ is the initial number of vehicles in segment $j$. $e_{i,j}$ indicates whether the vehicle is currently inside the roundabout; otherwise, $e_{i,j}=0$. We denote $\varPsi_{j}$ to be the auxiliary variable for absolute density deviation:
\begin{equation} \label{eq:ws}
\varPsi_{j} \geq \rho_{j} - \bar{\rho},\quad \varPsi_{j} \geq -(\rho_{j} - \bar{\rho}),\quad \varPsi_{j} \geq 0
\end{equation}




\subsubsection{The Multi-Scale Optimal Problem}
\par Finally, the multi-scale cost function is expressed as follows:
\begin{equation}
\begin{aligned}
&\min_{C, Y_{\Delta d,n}, \chi_{in}} \quad \alpha_1 \sum_{i=1}^{N} B (Y_{\Delta d,n})^2 + \alpha_2 \sum_{i=1}^{N} M (\chi_{in})^2 \\
&+ \alpha_3 \frac{1}{N} \sum_{i=1}^{N} c_{in} t_{i}^{\text{travel}} + \alpha_4 \sum_{s=1}^{V} \varPsi_{j}
\label{eq:objective_function}
\end{aligned}
\end{equation}
where $B$ and $M$ are constant weights. $\alpha_1$, $\alpha_2$, and $\alpha_3$ are scalar weighting coefficients for the multi-scale objectives. 
The constraints of system are followed by \cite{ref19} and the constraints of travel time are cited as \cite{ref17}. The classical branch-and-bound algorithm is applied to solve the optimization problem in Equation~(16), which determines the entering sequence of multiple vehicles at different conflict points in the roundabout.

\par Therefore, the optimal control algorithm of roundabout is given in Algorithm 1:
\begin{algorithm}[t]
\caption{The Optimal Roundabout Control Algorithm} 
\KwIn{Reference state $\{s^{\text{ref}}(0)\}$, road/roundabout map, prediction horizon $N_p$, vehicle constraints, $K$}
\KwOut{Real-time control inputs $\{a_i(k),\delta_i(k)\}$ and entering sequence $C(k)$ when applicable}

\textbf{Initialization:}\;
Read reference states $\{s^{\text{ref}}(0)\}$, initialize assumed sequences $C(0)$\; 
Set $t \gets 0$ and termination flags\;

\While{$t < K$ and $\text{not terminated}$}{ 
    \textbf{Step 1: Communication between vehicles}\;
    Determine whether the surrounding states are successfully collected based on $\delta(t)$\;
    Collect local/neighbor states and update the system state $\hat{s}(t+1)$ for the next time step $t+1$\;
    Assumed sequences $c_{i,n}$ from $i,n\in \mathcal{N}$\; 

    \textbf{Step 2: Optimal control of intersections}\;
    \If{vehicle $i$ is within sequence-trigger region of a merge point}{ 
        Formulate sequence matrix for vehicles near the merge point\;
        Build multi-objective optimization problem with: vehicle state/input limits, distance/velocity indicators, vehicle travel time, traffic density, sequence constraints\; 
        Solve optimization $\Rightarrow$ obtain optimal sequence $C^*(t)$\;
        Broadcast $C^*(t)$ to affected vehicles in the merge point\;
    } 

    \textbf{Step 3: Optimal control of vehicles}\;
    Calculate the TTC to determine the preceding and following vehicles of vehicle $i$\; 
    Build DMPC optimization problem with: vehicle dynamics, state/input limits, safety constraints with communication delay, sequence $C^*(t:t+N_p-1)$, control cost\;
    Solve optimization $\Rightarrow$ obtain control input $u_i^*(t:t+N_p-1)$\;
    Apply control input $u_i^*(t)$ to calculate the acceleration $a_i(t)$ and the steering $\delta_i(t)$\; 
    Update actual state $\hat{s}_i(t+1)$ using vehicle model\;

    \textbf{Step 4: Termination Check \& Time Advance}\;
    Check termination conditions\;
    $t \gets t + 1$\;
}

\Return{Executed control commands and logged sequences}
\end{algorithm}

\section{PERFORMANCE EVALUATION}

\subsection{Simulation Settings}
\par This section presents a comparison of the simulation results for three control algorithms in a roundabout environment to evaluate the effectiveness of the proposed algorithm: (1) a traditional model predictive control (MPC) strategy \cite{ref18}, (2) a two-layer control algorithm \cite{ref15}, and (3) the proposed algorithm. The basic simulation parameters are configured as follows: $L=60$~m ($\forall j \in \{1,2,3\}$), $\varDelta d^*=10$~m, maximum velocity $v_{\max}=20$~m/s, minimum velocity $v_{\min}=0$~m/s. The maximum acceleration for all vehicles is set to $u_{\max}=5$~m/s$^2$, and the minimum acceleration is $u_{\min}=-5$~m/s$^2$. 
The reactive time of vehicle is $1.8$~s, It is assumed that all vehicles in the mixed traffic flow maintain average driving behavior. According to the three-entry three-exit roundabout scenario illustrated in Fig. 1, the sampling step $\Delta t$ is set to $0.1\ \text{s}$, and the total simulation time is configured as $2000\ \text{s}$. The following metrics are selected to evaluate system efficiency: average energy consumption and travel time, which together reflect overall system performance.

\subsection{Experimental Results}

\subsubsection{Experiment~1}
\par The incoming traffic flow is generated according to a Poisson process, with all vehicle flow rates set to 396~veh/h and exit points assigned randomly. 200 vehicles randomly distributed across three intersections entered the roundabout following a Poisson distribution. We evaluated the performance of three control algorithms across four scenarios under different CAV penetration rates. 
The behavior of human-driven vehicles (HDVs) followed the same logic in each case, with their speed maintained at a constant 15 m/s. HDV movements triggered the lane-changing maneuvers of CAVs. Table I, Table II and Table III respectively present the comparative simulation results of multiple performance metrics across three intersections.

\par In Experiment~1, for different CAV penetration rates, the penetration rate was adjusted from 0.2 to 0.8 with an increment of 0.2 while keeping the initial states unchanged, as well as maintaining all other parameter settings. When the CAV penetration rate is low, the number of CAVs in the system is small, and the performance of the three control algorithms shows little difference. The MPC algorithm exhibited slightly better performance than the two-layer control algorithm when the CAV penetration rate was low. However, as the number of CAVs increased, the two-layer control algorithm steadily improved the system performance by scheduling the vehicle passing sequence. With the increase in the CAV penetration rate, the performance of the control algorithm proposed in this paper is significantly superior to that of the comparison algorithms. At penetration rates of 0.6 and 0.8, for Intersection A, the performance in terms of average vehicle travel time and energy consumption is improved by 15\% and 16\%, respectively. For Intersection B, the corresponding performance improvements are 40\% and 27\%, while for Intersection C, the improvements reach 41\% and 47\%. Through the scheduling and motion control of a certain scale of CAVs, the overall traffic performance of the system can be effectively guaranteed.

\begin{figure*}[!t]
    \centering
    \subcaptionbox{The performance comparison of PET under CAV penetration rate 0.6\label{subfig:travel}}{%
        \includegraphics[width=0.235\textwidth]{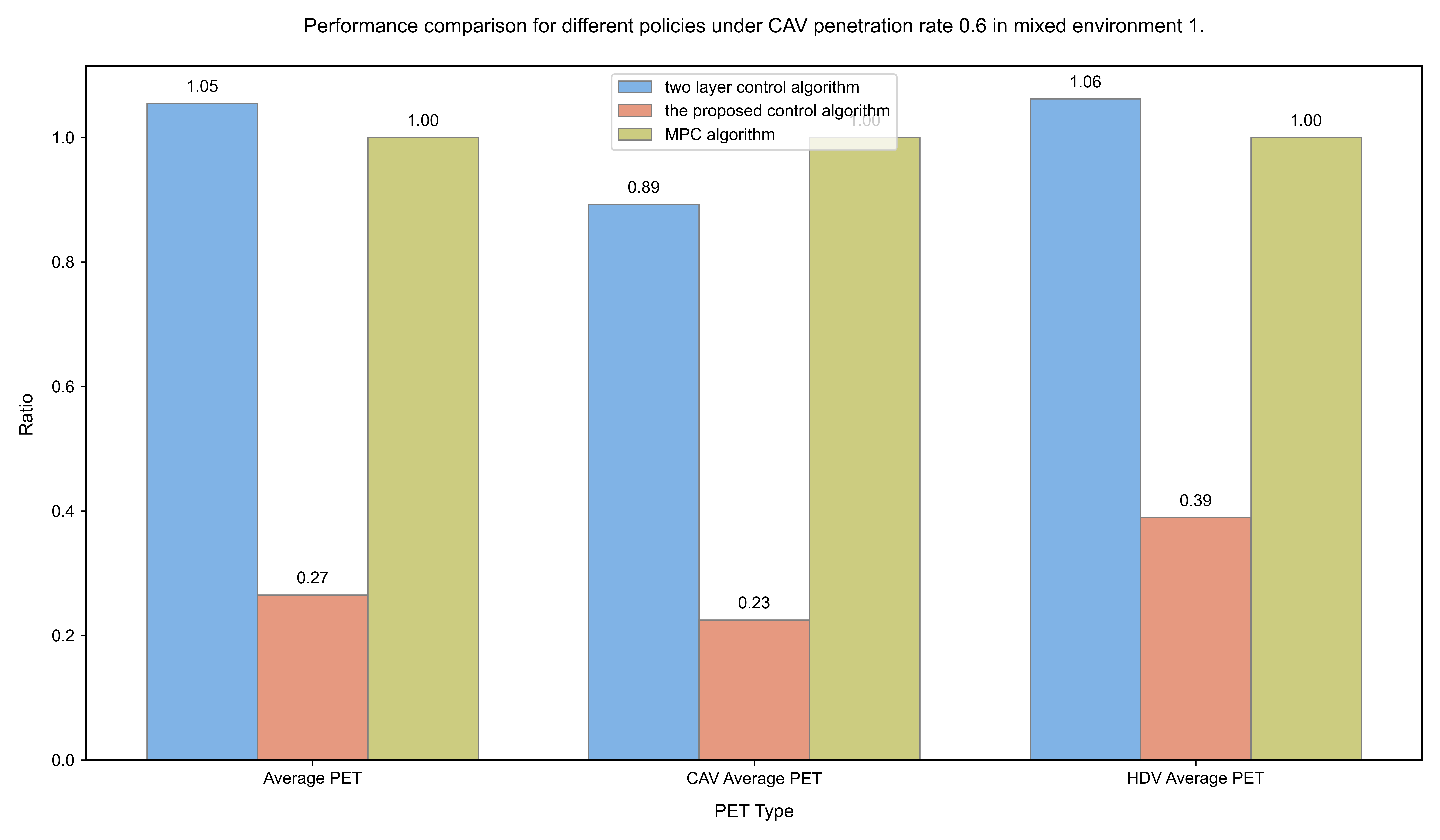}
    }
    \hfill 
    \subcaptionbox{The performance comparison of conflict ratio under CAV penetration rate 0.6\label{subfig:energy}}{%
        \includegraphics[width=0.235\textwidth]{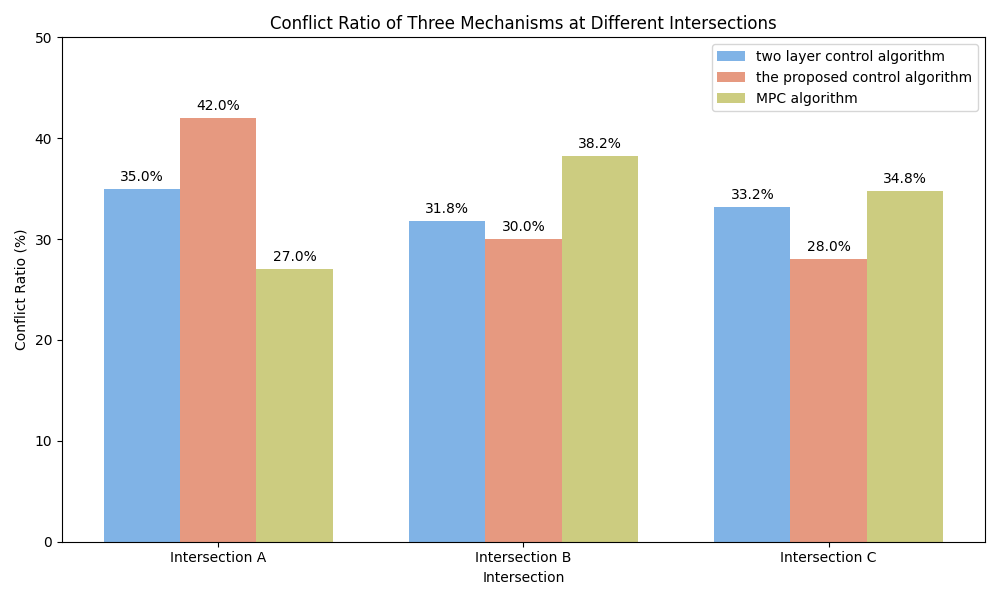}
    }
    \hfill
    \subcaptionbox{The performance comparison of PET under CAV penetration rate 0.8\label{subfig:obj01}}{%
        \includegraphics[width=0.235\textwidth]{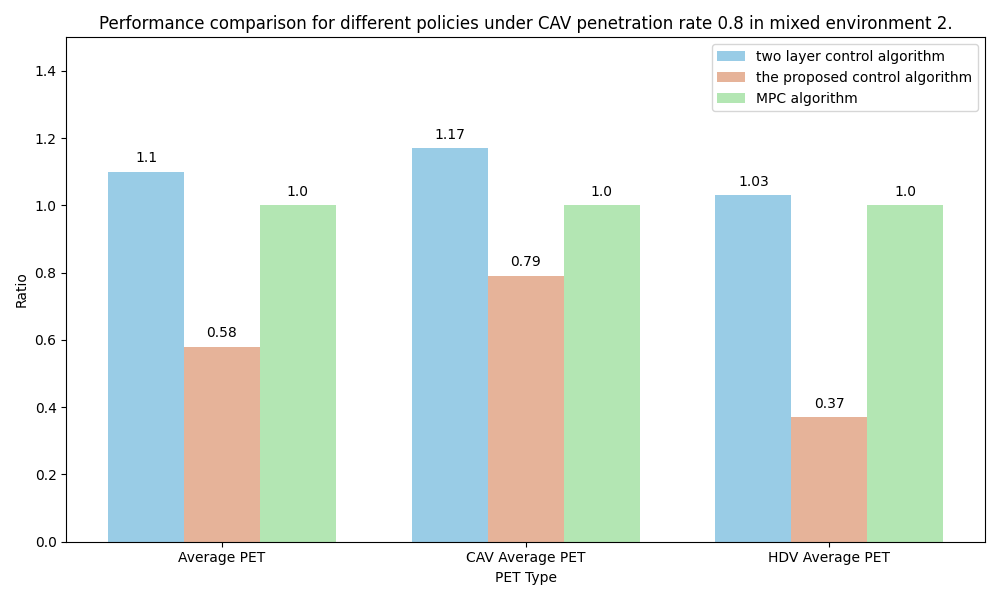}
    }
    \hfill
    \subcaptionbox{The performance comparison of conflict ratio under CAV penetration rate 0.8\label{subfig:obj02}}{%
        \includegraphics[width=0.235\textwidth]{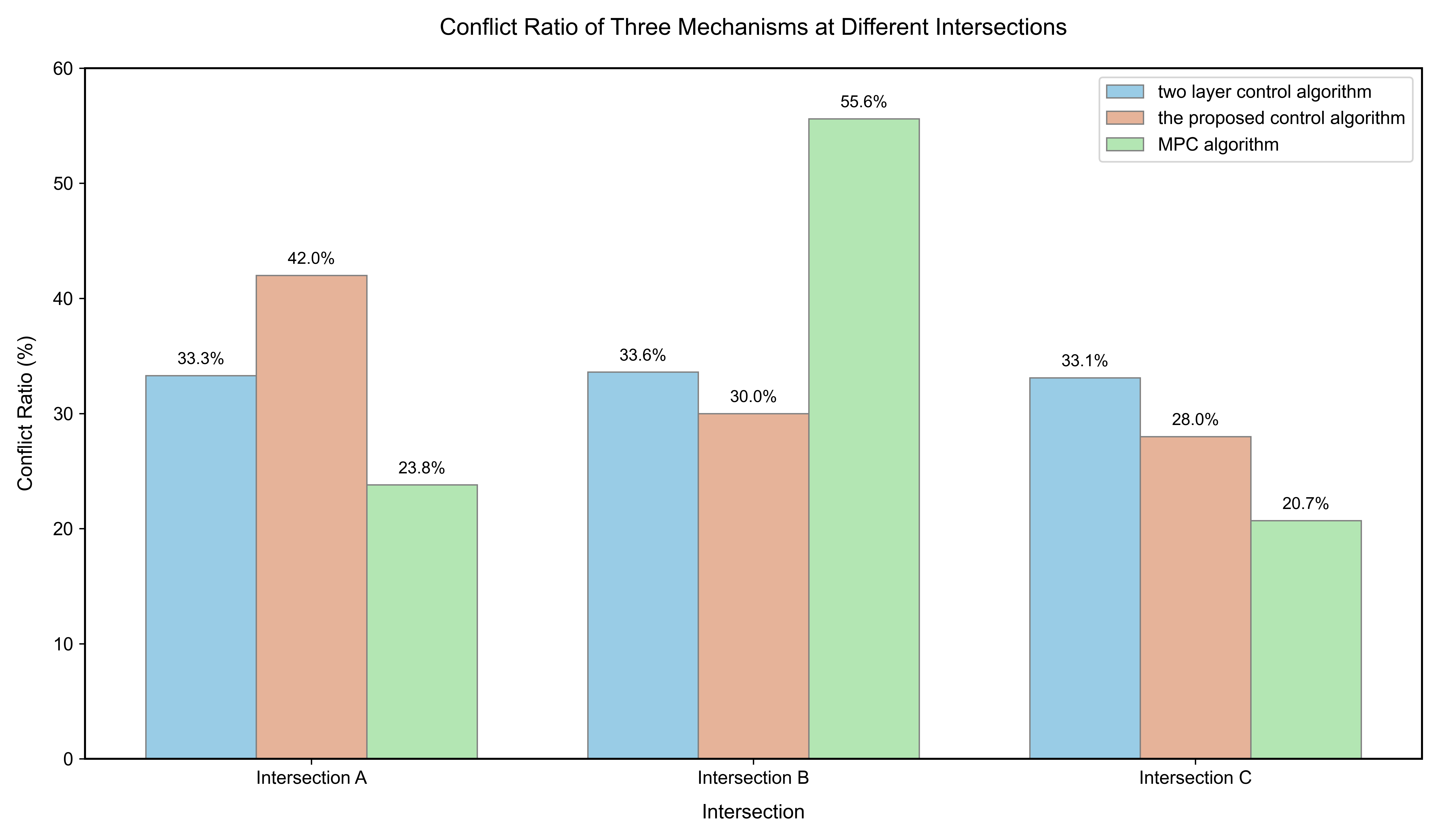}
    }
    
    \caption{The performance comparison of PET and conflict radio with different algorithms}
    \label{fig:all_metrics}
\end{figure*}

\begin{table*}[htbp]
\centering
\footnotesize 
\caption{Performance comparison of different control algorithms for Intersection A under different CAV penetration rates}
\begin{tabular}{|l|*{12}{c|}} 
\hline 
\diagbox[width=10em, height=4em]{\textbf{Attribute}}{\textbf{Intersection A}} 
    & \multicolumn{3}{c|}{0.2} & \multicolumn{3}{c|}{0.4} & \multicolumn{3}{c|}{0.6} & \multicolumn{3}{c|}{0.8} \\
\cline{2-13} 
& M1 & M2 & M3 & M1 & M2 & M3 & M1 & M2 & M3 & M1 & M2 & M3 \\
\hline 
Ave. Travel Time & 10.12 & 10.74 & 10.91 & 19.79 & 15.72 & 17.77 & 23.17 & 19.60 & 23.39 & 33.38 & 26.16 & 49.40 \\
\hline 
Ave. Energy & 14.05 & 14.92 & 15.15 & 27.48 & 21.84 & 24.68 & 32.19 & 27.22 & 32.49 & 46.37 & 36.33 & 68.61 \\
\hline 
Ave. obj($\lambda=0.1$) & 28.10 & 29.84 & 30.29 & 54.97 & 43.68 & 49.35 & 64.37 & 54.45 & 64.99 & 92.74 & 72.66 & 137.23 \\
\hline 
Ave. obj($\lambda=0.2$) & 45.66 & 48.49 & 49.22 & 89.32 & 70.98 & 80.19 & 104.60 & 88.47 & 105.60 & 150.69 & 118.06 & 222.98 \\
\hline 
\end{tabular}
\vspace{1mm}
\par \footnotesize Note: M1 corresponds to the two-layer control algorithm \cite{ref15}, M2 corresponds to the control algorithm proposed in this paper, and M3 corresponds to the MPC algorithm \cite{ref18}. $\text{Average obj}=\eta \cdot \text{Average Travel time}+\text{Average Energy}$. $\text{Average Energy}=\text{Average Travel time}+\sum_{k=0}^{K} \frac{1}{2} u_{i}^{2}(k)\, \Delta t$.
\par $\eta$ denotes the weight coefficient in the function and $\eta =\frac{\lambda \cdot \max \left\{ a_{\max}^{2},a_{\min}^{2} \right\}}{2\left( 1-\lambda \right)}$.
\end{table*}
\begin{table*}[htbp]
\centering
\footnotesize 
\caption{Performance comparison of different control algorithms for Intersection B under different CAV penetration rates}
\begin{tabular}{|l|*{12}{c|}}
\hline 
\diagbox[width=10em, height=4em]{\textbf{Attribute}}{\textbf{Intersection B}} 
    & \multicolumn{3}{c|}{0.2} & \multicolumn{3}{c|}{0.4} & \multicolumn{3}{c|}{0.6} & \multicolumn{3}{c|}{0.8} \\
\cline{2-13} 
& M1 & M2 & M3 & M1 & M2 & M3 & M1 & M2 & M3 & M1 & M2 & M3 \\
\hline 
Ave. Travel Time & 12.88 & 8.94 & 13.21 & 29.45 & 18.64 & 36.18 & 52.49 & 31.31 & 42.82 & 60.61 & 40.81 & 64.89 \\
\hline 
Ave. Energy & 17.90 & 12.42 & 18.35 & 40.90 & 25.89 & 50.25 & 72.90 & 43.49 & 59.48 & 84.18 & 56.68 & 90.14 \\
\hline 
Ave. obj($\lambda=0.1$) & 35.79 & 24.85 & 36.70 & 81.80 & 51.78 & 100.51 & 145.80 & 86.99 & 118.95 & 168.36 & 113.36 & 180.28 \\
\hline 
Ave. obj($\lambda=0.2$) & 58.16 & 40.37 & 59.63 & 132.92 & 84.13 & 163.31 & 236.92 & 141.34 & 193.29 & 273.58 & 184.20 & 292.93 \\
\hline 
\end{tabular}
\vspace{1mm}
\par \footnotesize Note: M1 corresponds to the two-layer control algorithm \cite{ref15}, M2 corresponds to the control algorithm proposed in this paper, and M3 corresponds to the MPC algorithm \cite{ref18}. $\text{Average obj}=\eta \cdot \text{Average Travel time}+\text{Average Energy}$. $\text{Average Energy}=\text{Average Travel time}+\sum_{k=0}^{K} \frac{1}{2} u_{i}^{2}(k)\, \Delta t$.
\par $\eta$ denotes the weight coefficient in the function and $\eta =\frac{\lambda \cdot \max \left\{ a_{\max}^{2},a_{\min}^{2} \right\}}{2\left( 1-\lambda \right)}$.
\end{table*}
\begin{table*}[htbp]
\centering
\footnotesize 
\caption{Performance comparison of different control algorithms for Intersection C under different CAV penetration rates}
\begin{tabular}{|l|*{12}{c|}}
\hline 
\diagbox[width=10em, height=4em]{\textbf{Attribute}}{\textbf{Intersection C}} 
    & \multicolumn{3}{c|}{0.2} & \multicolumn{3}{c|}{0.4} & \multicolumn{3}{c|}{0.6} & \multicolumn{3}{c|}{0.8} \\
\cline{2-13} 
& M1 & M2 & M3 & M1 & M2 & M3 & M1 & M2 & M3 & M1 & M2 & M3 \\
\hline 
Ave. Travel Time & 19.03 & 14.78 & 24.03 & 33.79 & 25.65 & 43.73 & 55.42 & 32.40 & 61.42 & 68.25 & 38.76 & 70.05 \\
\hline 
Ave. Energy & 26.43 & 20.53 & 33.38 & 46.94 & 35.63 & 60.74 & 76.97 & 45.01 & 85.31 & 94.81 & 53.84 & 97.30 \\
\hline 
Ave. obj($\lambda=0.1$) & 52.86 & 41.07 & 66.77 & 93.88 & 71.26 & 121.47 & 153.95 & 90.01 & 170.62 & 189.61 & 107.67 & 194.59 \\
\hline 
Ave. obj($\lambda=0.2$) & 85.89 & 66.73 & 108.49 & 152.55 & 115.80 & 197.39 & 250.15 & 146.26 & 277.24 & 308.10 & 174.96 & 316.19 \\
\hline 
\end{tabular}
\vspace{1mm}
\par \footnotesize Note: M1 corresponds to the two-layer control algorithm \cite{ref15}, M2 corresponds to the control algorithm proposed in this paper, and M3 corresponds to the MPC algorithm \cite{ref18}. $\text{Average obj}=\eta \cdot \text{Average Travel time}+\text{Average Energy}$. $\text{Average Energy}=\text{Average Travel time}+\sum_{k=0}^{K} \frac{1}{2} u_{i}^{2}(k)\, \Delta t$.
\par $\eta$ denotes the weight coefficient in the function and $\eta =\frac{\lambda \cdot \max \left\{ a_{\max}^{2},a_{\min}^{2} \right\}}{2\left( 1-\lambda \right)}$.
\end{table*}

\par In the simulation experiments, we set the Post-Encroachment Time (PET) as the index for quantifying the collision risk of vehicles within the roundabout, which is derived by measuring the time difference between two vehicles in the conflict zone \cite{ref29}.
Based on this, under the scenarios with 60\% and 80\% CAV penetration rates, we respectively calculate various PET metrics at the three intersections in the two scenarios, with the specific results presented in Fig. 3. Fig. 3(a) and Fig. 3(c) respectively represent the average PET of the entire roundabout system under the two scenarios, and separately quantify the PET values corresponding to CAVs and HDVs. Fig. 3(b) and Fig. 3(d) count the conflict probability of each intersection in the two scenarios. The results of the simulation experiments are illustrated in Fig. 3, where we set the average index values of the Model Predictive Control (MPC) algorithm \cite{ref18} as the baseline and normalize them to 1.
The results of the other two cases are presented as ratios relative to this baseline.

\par In Experiment~1, the performance of the MPC algorithm and the two-layer control algorithm is relatively close, while the control algorithm proposed in this paper significantly reduces the conflict probability. When the CAV penetration rate reaches 0.6, the collision conflicts of CAVs are reduced to 0.23, and the vehicle safe driving performance is improved by 70\%. Although the mechanism proposed in this paper results in a relatively high conflict rate at Intersection A, it still ensures the safe driving of vehicles. With the increase in the number of CAVs, the proposed mechanism is superior in terms of overall performance. Although the number of collision conflicts increases, the mechanism adaptively responds to the changes in vehicles within the system through the sequential scheduling of vehicles, and the performance is still improved by 42\%. Meanwhile, it balances the conflict rate of each intersection, laying a foundation for further improving the roundabout performance. This superiority is mainly attributed to the fact that the proposed algorithm fully considers the system traffic flow when implementing sequential control of vehicles, and adaptively adjusts the vehicle entry speed in a dynamic manner to avoid significant congestion in the roundabout as the number of CAVs increases. The aforementioned subfigures indicate that the sequential control and motion control of vehicles can significantly improve roundabout performance, reduce conflict probability remarkably, and ensure roundabout efficiency in a stable manner.

\subsubsection{Experiment~2}
\par The traffic inflow rates at merge points $O_1$, $O_2$, and $O_3$ are set to 108~veh/h, 540~veh/h, and 540~veh/h, respectively, while all other parameters remain unchanged. Additionally, the vehicle arrival rate at all exit nodes is increased to 576~veh/h, representing a 2.5-fold increase in traffic demand compared to the previous case, in order to simulate congested conditions. The remaining parameters are kept constant to facilitate comparison of the system performance of the three algorithms under heavy traffic demand.

\par In Experiment~2, the total number of vehicles is increased to simulate the performance of the three algorithms under a dense roundabout scenario, and the results are summarized in Table~IV. Different traffic demands are assigned to the three entries to emulate the influence of varying flow rates within the three conflict zones under each of the three algorithms. At Intersection A, the control algorithm proposed in this paper achieves a 10\% improvement compared with the two-layer control algorithm and a 7\% improvement compared with the MPC algorithm. At Intersection B, with the increase in the number of vehicles, the performance improvements are 13\% and 2\%, respectively. At Intersection C, the performance improvements reach 16\% and 1.2\%, respectively. When the algorithm proposed in this paper is adopted, not only is the flow in the current conflict zone considered, but the flows in the other two conflict zones are also jointly taken into account. This ensures both effective vehicle passage and optimal overall system performance. Even when a large number of vehicles exist in the roundabout, the control algorithm proposed in this paper can always maintain good performance. In contrast, the two-layer algorithm can only respond to partial flow variations within the small-scale self-organizing communication region. Moreover, due to obstacle-avoidance requirements, the waiting time within conflict zones increases significantly. Although both algorithms show limited differences in system-level performance indicators within the same simulation duration, the total number of vehicles successfully passing through the roundabout drops noticeably. 

\begin{table*}[!t]
  \centering
  \caption{Performance comparison of different control algorithms for three intersections under the heavy environment}
  \label{tab:combined_4x4}
  \begin{tabular}{|c|c|c|c|}
    \hline
    \textbf{Intersection} & \textbf{two-layer control algorithm\cite{ref15}} & 
    \textbf{the proposed control algorithm} & 
    \textbf{MPC algorithm\cite{ref18}} \\ \hline

    \multicolumn{4}{|c|}{\textbf{Intersection A}} \\ \hline
    Average Travel time & 55.8437 & 50.2507 & 54.2085 \\ \hline
    Average Energy      & 77.5668 & 69.7982 & 75.2955 \\ \hline
    Average obj($\lambda=0.1$)        & 155.1337   & 139.5965   & 150.5911 \\ \hline
    Average obj($\lambda=0.2$)        & 252.0783    & 226.8317    & 244.6969 \\ \hline

    \multicolumn{4}{|c|}{\textbf{Intersection B}} \\ \hline
    Average Travel time & 214.7484 & 185.8144 & 189.9260 \\ \hline
    Average Energy      & 298.2855 & 258.0962 & 263.8073 \\ \hline
    Average obj($\lambda=0.1$)        & 596.5710    & 516.1925    & 527.6146 \\ \hline
    Average obj($\lambda=0.2$)        & 969.3742    & 838.7663    & 857.3262 \\ \hline

    \multicolumn{4}{|c|}{\textbf{Intersection C}} \\ \hline
    Average Travel time & 452.1850 & 378.8318 & 383.7575 \\ \hline
    Average Energy      & 628.0850 & 526.1973 & 533.0391 \\ \hline
    Average obj($\lambda=0.1$)        & 1256.1701    & 1052.3947    & 1066.0783 \\ \hline
    Average obj($\lambda=0.2$)        & 2041.1633    & 1710.0466    & 1732.2813 \\ \hline
  \end{tabular}
  \vspace{1mm}
\par \footnotesize Note: $\text{Average obj}=\eta *\text{Average Travel time}+\text{Average Energy}$. $\text{Average Energy}={\text{Average Travel time}}+\sum_{k=0}^{K} \frac{1}{2} u_{i}^{2}(k)\, \Delta t$.
\par $\eta $ denotes the weight coefficient in the function and $\eta =\frac{\lambda \cdot \max \left\{ a_{\max}^{2},a_{\min}^{2} \right\}}{2\left( 1-\lambda \right)}$.
\end{table*}

\section{CONCLUSION}
\par This paper investigates a roundabout control algorithm that accounts for communication latency and inaccuracies in interactive information. First, by considering the impact of communication performance on control, a vehicle motion model is proposed, and the motion influences of preceding and following vehicles on the current vehicle are determined based on time-to-collision (TTC). Next, based on distributed model predictive control (DMPC), the optimal distributed control inputs are derived by integrating vehicle state information obtained through communication interactions, thereby guiding vehicles to drive safely and maintain a safe distance from preceding and following vehicles at all times. Then, a multi-scale optimization objective is constructed to formulate a sequential control optimization problem, which guides vehicles to safely enter the signal-free roundabout while ensuring traffic efficiency. Finally, the effectiveness of the proposed control algorithm is verified through simulation experiments under various traffic scenarios.


\begin{thebibliography}{00}

\bibitem{ref1} X. Wang, J. Liu, T. Qiu, C. Mu, C. Chen, and P. Zhou, A real-time collision prediction mechanism with deep learning for intelligent transportation system, IEEE Transactions on Vehicular Technology, 69(9): 9497–9508, 2020.
\bibitem{ref2} D. L. Fisher, M. Lohrenz, D. Moore, E. D. Nadler, and J. K. Pollard, Humans and intelligent vehicles: The hope,the help,and the harm, IEEE Transactions on Intelligent Vehicles, 1(1): 56–67, 2016.
\bibitem{ref3} Y.Wang, Z.Liu, Z.Zuo, Z.Li, L.Wang, and X.Luo, Trajectory planning and safety assessment of autonomous vehicles based on motion prediction and model predictive control, IEEE Transactions on Vehicular Technology, 68(9): 8546–8556, 2019.
\bibitem{ref4} S. Mandavilli, M. J. Rys, and E. R. Russell, Environmental impact of modern roundabouts, International Journal of Industrial Ergonomics, 38(2): 135–142, 2008.
\bibitem{ref5} Deluka Tibljaˇ s, A., Giuffr`e, T., Surdonja, S. and Trubia, S., Introduction of Autonomous Vehicles: Roundabouts Design and Safety Performance Evaluation, Sustainability, 10(4):1060, 2018. 
\bibitem{ref6} A. Flannery and T. Datta, Operational performance measures of american roundabouts, Transportation Research Record, 1572(1): 68–75, 1997. 
\bibitem{ref7} A. Flannery, L. Elefteriadou, P. Koza, and J. McFadden, Safety, delay, and capacity of single-lane roundabouts in the united states, Transportation Research Record, 1646(1): 63–70, 1998. 
\bibitem{ref8} H. M. Al-Madani, Dynamic vehicular delay comparison between a police-controlled roundabout and a traffic signal, Transportation Research Part A: Policy and Practice, 37(8): 681–688, 2003. 
\bibitem{ref9} V. P. Sisiopiku and H.-U. Oh, Evaluation of roundabout performance using sidra, Journal of Transportation Engineering, 127(2): 143–150, 2001. 
\bibitem{ref10} S. Mandavilli, M. J. Rys, and E. R. Russell, Environmental impact of modern roundabouts, International Journal of Industrial Ergonomics, 38(2): 135–142, 2008. 
\bibitem{ref11} J. Rios-Torres and A. A. Malikopoulos, A survey on the coordination of connected and automated vehicles at intersections and merging at highway on-ramps, IEEE Transactions on Intelligent Transportation Systems, 18(5): 1066–1077, 2016.
\bibitem{ref12} Zohdy, I.H., Kamalanathsharma, R.K. and Rakha, H., Intersection management for autonomous vehicles using iCACC, In 2021 15th International IEEE Conference on Intelligent Transportation Systems, pages 1109-1114, IEEE, 2021 . 
\bibitem{ref13} Bayar, B., Sajadi-Alamdari, S.A., Viti, F. and Voos, H., Impact of different spacing policies for adaptive cruise control on traffic and energy consumption of electric vehicles, In 2016 24th Mediterranean Conference on Control and Automation (MED), pages 1349-1354, IEEE, 2016.
\bibitem{ref14} O. Rashed and R. Imam, A functional and operational comparison between signalized and unsignalized roundabouts, International Journal of Engineering Research and Technology, 13(6): 1448, 2020.
\bibitem{ref15} X. Liu, Q. Chen, P. Hang, L. Xiong and J. Sun, A Self-organizing Cooperative Control Framework for Connected Automated Vehicles at Unsignalized Roundabouts, In 2024 IEEE 27th International Conference on Intelligent Transportation Systems (ITSC), pages 358-364, IEEE, 2024.
\bibitem{ref16} H. Jafarzadeh and C. Fleming, Gaussian Process-based Model Predictive Controller for Connected Vehicles with Uncertain Wireless Channel, In 2021 IEEE International Intelligent Transportation Systems Conference (ITSC), pages 3515-3520, IEEE, 2021.
\bibitem{ref17} S. V. Patil, K. Hashimoto and M. Kishida, A Robust Traffic Flow Control Using Connected Vehicle Technology: Signal Spatio-Temporal Logic-Based Approach, IEEE Transactions on Intelligent Transportation Systems, 25(12): 19658-19674, 2024. 
\bibitem{ref18} A. Bozzi, S. Graffione, R. Sacile and E. Zero, Cooperative Roundabout Navigation: Distributed Predictive Control with Virtual Platooning in Mixed Traffic, In 2024 IEEE 20th International Conference on Automation Science and Engineering (CASE), pages 546-551, IEEE, 2024.
\bibitem{ref19} S. Li, Y. Zhou, X. Ye, J. Jiang and M. Wang, Sequencing-Enabled Hierarchical Cooperative On-Ramp Merging Control for Connected and Automated Vehicles, In 2023 IEEE 26th International Conference on Intelligent Transportation Systems (ITSC), pages 5146-5153, IEEE, 2023.
\bibitem{ref20} Y. Zhong, G. Lyu, X. He, Y. Zhang and S. S. Ge, Distributed Active Fault-Tolerant Cooperative Control for Multiagent Systems With Communication Delays and External Disturbances, IEEE Transactions on Cybernetics, 53(7): 4642-4652, 2023.
\bibitem{ref21} C. Deng, L. Xu, T. Yang, D. Yue and T. Chai, Distributed Cooperative Optimization for Nonlinear Heterogeneous MASs Under Intermittent Communication, IEEE Transactions on Automatic Control,  69(4): 2737-2744, 2024.
\bibitem{ref22} S. Sudhakara, D. Kartik, R. Jain and A. Nayyar, Optimal Communication and Control Strategies in a Cooperative Multiagent MDP Problem, IEEE Transactions on Automatic Control, 69(10): 6959-6966, 2024.
\bibitem{ref23} X. Duan and Y. Zhao and D. Tian and J. Zhou, L. Ma and L. Zhang, Joint Communication and Control Optimization of a UAV-Assisted Multi-Vehicle Platooning System in Uncertain Communication Environment, IEEE Transactions on Vehicular Technology, 73(3): 3177-3190, 2023.
\bibitem{ref24} K. Xu, C. G. Cassandras and W. Xiao, Decentralized Time and Energy-Optimal Control of Connected and Automated Vehicles in a Roundabout, In 2021 IEEE International Intelligent Transportation Systems Conference (ITSC), pages 681-686, IEEE, 2021.
\bibitem{ref25} Y. Chen, C. G. Cassandras and K. Xu, Optimal Sequencing and Motion Control in a Roundabout with Safety Guarantees, In 2024 IEEE 63rd Conference on Decision and Control (CDC), pages 5699-5704, IEEE, 2024.
\bibitem{ref26} O. Pauca and C. F. Caruntu, Travel Time Minimization at Roundabouts for Connected and Automated Vehicles, In 2020 25th IEEE International Conference on Emerging Technologies and Factory Automation (ETFA), pages 905-910, IEEE, 2020.
\bibitem{ref27} A. Haraldsen, M. S. Wiig, A. D. Ames, K. Y. Pattersen. Safety-Critical Control of Nonholonomic Vehicles in Dynamic Environments Using Velocity Obstacles, In 2024 American Control Conference (ACC), pages 3152-3159, IEEE, 2024.
\bibitem{ref28}
Y. Chen, C. G. Cassandras and K. Xu, Optimal Sequencing and Motion Control in a Roundabout with Safety Guarantees, In 2024 IEEE 63rd Conference on Decision and Control (CDC), pages 5699-5704, IEEE, 2024.
\bibitem{ref29}
M. Paul and I. Ghosh, Post encroachment time threshold identification for right-turn related crashes at unsignalized intersections on intercity highways under mixed traffic, International journal of injury control and safety promotion, 27(2): 121–135, 2020.
\end{thebibliography}
\end{document}